\documentclass[final,3p,times]{elsarticle}
\usepackage{arydshln}
\usepackage{geometry,latexsym,amssymb,amsmath,framed,amsbsy,mathtools,wasysym}
\usepackage{caption,subcaption}
\usepackage{tikz}
   \usetikzlibrary{3d}
   \usetikzlibrary{calc}

\definecolor{mygreen}{HTML}{23b626}
\usepackage{verbatim}
\usepackage{fancyvrb,color,xcolor}
\usepackage{listings}
\biboptions{sort&compress}

\newcounter{bla}

\journal{Computer Physics Communications}

\def\uu {\mathbf{u}}
\def\xx {\mathbf{x}}

\newcommand{\pluss}[1]{\tikz{\draw[#1,line width=0.6pt] (0,0.075) -- (0.15,0.075) (0.075,0) -- (0.075,0.15);}}
\newcommand{\cross}[1]{\tikz{\draw[#1,line width=0.6pt] (0,0) -- (0.15,0.15) (0,0.15) -- (0.15,0);}}
\newcommand{\starr}[1]{\tikz{\draw[#1,line width=0.6pt] (0,0) -- (0.15,0.15) (0,0.15) -- (0.15,0) (0,0.075) -- (0.15,0.075) (0.075,0) -- (0.075,0.15);}}

\newcommand{\osqr}[1]{\tikz{\draw[#1,line width=0.6pt] (0,0) -- (0.15,0) -- (0.15,0.15) -- (0,0.15) -- cycle;}}

\newcommand{\ocir}[1]{\tikz{\draw[#1,line width=0.6pt] (0,0) circle (0.075);}}

\newcommand{\fttru}[1]{\tikz{\filldraw[#1,line width=0.6pt] (0,0) -- (60:0.17) -- (0.173,0) -- cycle;}}
\newcommand{\ottru}[1]{\tikz{\draw[#1,line width=0.6pt] (0,0) -- (60:0.173) -- (0.173,0) -- cycle;}}
\newcommand{\fttrd}[1]{\tikz{\filldraw[#1,line width=0.6pt] (0,0) -- (300:0.173) -- (0.173,0) -- cycle;}}
\newcommand{\ottrd}[1]{\tikz{\draw[#1,line width=0.6pt] (0,0) -- (300:0.173) -- (0.173,0) -- cycle;}}
\newcommand{\frhom}[1]{\tikz{\filldraw[#1,line width=0.6pt] (0,0) -- (0.0866,0.0866) -- (0.173,0) -- (0.0866,-0.0866) -- cycle;}}
\newcommand{\orhom}[1]{\tikz{\draw[#1,line width=0.6pt] (0,0) -- (0.0866,0.0866) -- (0.173,0) -- (0.0866,-0.0866) -- cycle;}}

\begin{document}

\begin{frontmatter}



%\title{A partitioned global address space (PGAS) programming based 
%algorithm for particle tracking in turbulence simulations at
%extreme problem sizes} 
\title{A highly scalable particle tracking algorithm using 
partitioned global address space (PGAS) 
programming for extreme-scale turbulence simulations}


\author[a,b]{Dhawal Buaria\corref{author}}
\author[b,c]{P.K. Yeung}

\cortext[author] {Corresponding author. \textit{E-mail address:} 
dhawal.buaria@ds.mpg.de}
\address[a]{Max-Planck Institute for Dynamics and Self-Organization, 
D-37077 G\"ottingen, Germany}
\address[b]{School of Aerospace Engineering, Georgia Institute of Technology, 
Atlanta, GA 30332, USA}
\address[c]{School of Mechanical Engineering, Georgia Institute of Technology, 
Atlanta, GA 30332, USA}

\begin{abstract}

A new parallel algorithm utilizing 
partitioned global address space (PGAS) programming model to
achieve high scalability is reported
for particle tracking in direct numerical simulations of turbulent flow.
The work is motivated by the desire to obtain Lagrangian information
necessary for the study of turbulent dispersion
at the largest problem sizes feasible on 
current and next-generation multi-petaflop supercomputers.
A large population of fluid particles is distributed among parallel processes
dynamically, based on instantaneous particle positions such that all of the
interpolation information needed for each particle is available either locally
on its host process or neighboring processes holding adjacent sub-domains
of the velocity field. With cubic splines as the preferred interpolation method, 
the new algorithm is designed to minimize the need for communication,
by transferring between adjacent processes only those spline coefficients
determined to be necessary for specific particles.
This transfer is implemented very efficiently
as a one-sided communication, using
Co-Array Fortran (CAF) features which facilitate small data movements
between different local partitions of a large global array.
The cost of monitoring transfer of particle
properties  between adjacent processes for particles
migrating across sub-domain boundaries is found to be small.
Detailed benchmarks are obtained on the Cray petascale supercomputer {\em Blue Waters} 
at the University of Illinois, Urbana-Champaign. 
For operations on the particles in a $8192^3$
simulation 
(0.55 trillion grid points) 
on 262,144 Cray XE6  cores, the new algorithm is found to be orders of magnitude faster
relative to a prior algorithm in which each particle
is tracked by the same parallel process at all times.
This large speedup reduces the additional cost of tracking
of order 300 million particles to just over 50\% of the cost of computing
the Eulerian velocity field at this scale.
Improving support of PGAS models on major compilers suggests
that this algorithm will be of wider applicability on most upcoming supercomputers.

\end{abstract}
 

\end{frontmatter}

\section{Introduction}

Many important problems in nature and engineering,
such as pollutant dispersion,
cloud physics, and the design of improved combustion devices, are
closely tied to the motion of discrete entities
in a continuous fluid medium in a state of
turbulent motion, characterized by disorderly fluctuations 
in time and three-dimensional (3D) space. 
Important applications include 
the role of pollutant dispersion
in  atmospheric air quality
\citep{pasquill},
the coalescence of water vapor droplets leading
to rain formation \citep{Shaw.2003}, 
and the mixing of chemical species 
in turbulent combustion \citep{pope.1985}.
Although effects of particle inertia \citep{Elghobashi,biferale.prx}
and molecular diffusion \citep{saffman1960,SH.1986}
are often present,
the predominant physical mechanism underlying these applications
is that of turbulent transport via
the motion of infinitesimal material fluid elements,
known as fluid particles (or
passive tracers), 
which are (from the continuum viewpoint) of zero size
and move with the local flow velocity.
Effectively, we adopt the Lagrangian viewpoint
of fluid mechanics \cite{MY.I, MY.II,SP.2013} from the
perspective of an observer moving with the flow. 
A number of review articles 
covering various aspects of this
broad subject are given by Refs.
\citep{pope1994, sawford.2001, yeung2002, TB09, SC09, bala.arfm}.

The trajectory of a fluid particle 
can be obtained by numerical integration of
its equation of motion
\begin{align}
\frac{d\xx^+(t)}{dt} = \uu^+(t) \ ,
\label{eqn:dxdt}
\end{align}
where $\xx^+(t)$ and $\uu^+(t)$ denote the 
instantaneous particle position and velocity
respectively. The fluid particle velocity is given by the
velocity of the fluid medium at the instantaneous particle position, i.e.
\begin{align}
\uu^+(t) = \uu (\xx^+(t), t) \ ,
\end{align}
where $\uu(\xx,t)$ represents the  so-called Eulerian velocity field
seen by an observer at fixed locations in space.
To follow the particle trajectories it is thus necessary
to first calculate the fluid velocity at a set
of fixed grid locations, and then to interpolate
for the particle velocity based on its instantaneous
position and the Eulerian velocity field at a set of
neighboring grid points. Since instantaneous velocities
are involved, the only reliable computational technique
to obtain the Eulerian information required is direct
numerical simulation (DNS), where the velocity field is
computed numerically according to the Navier-Stokes equations
expressing the  fundamental laws of conservation of mass and momentum.
The value of DNS as a research tool capable of providing
massive detail is well established \citep{MM98,IGK2009}.

Turbulence is one of the main science drivers
for high-performance computing \citep{Yokokawa.2002,moserSC}.
A general desire is to
reach Reynolds numbers as high as possible, so that the 
flow physics captured will bear greater resemblance to
that in flows encountered in practical applications. 
The Reynolds number ($Re$) is a non-dimensional parameter 
defined as
$Re ={\cal U} \ell/\nu$, where ${\cal U}$ is 
a measure of the velocity fluctuations,
$\ell$ is a characteristic length scale of large scales,
and $\nu$ is the kinematic viscosity. 
A large Reynolds number implies a wide range of scales
in both time and space, which in turn requires a large
number of time steps and grid points. 
Depending on the detailed definitions used and
scale resolution desired,
it can be estimated that the total computational effort for
a simulation over a given physical time period
increases with Reynolds number 
at %mark2 
least as strongly as $Re^3$
\citep{pope.book}.
Advances in computing power have enabled 
simulations in  the simplest geometries at
$8192^3$ \cite{YZS.2015} and
even $12288^3$ \cite{Ishihara2016} grid resolution.
On other hand, although the basics of particle tracking
are well established
\citep{YP.1988, Bala89, HDG.2007}, 
further challenges arise when tracking
a large number of fluid particles
at such problem sizes. 
In particular, since the solution domain is distributed
over multiple parallel processes, as each  fluid particle wanders
around, the identities of 
parallel processes directly involved in the calculation of its
interpolated velocity
also evolve. Conversely, each parallel
process is called upon to contribute to the interpolated velocities
of a constantly-evolving  collection of fluid particles at each time step.
This can lead to a very inefficient communication pattern,
with adverse effects on performance and scalability.

In this paper our ultimate goal is to address the issues
associated with particle tracking at extreme problem sizes
within a massively parallel programming model, with
emphasis on challenges that may not arise or be evident
at smaller problem sizes.
The first decision in the design of any parallel particle tracking
algorithm is how the particles are divided among the
parallel processes   used for the Eulerian DNS.
One basic strategy is to assign to each process the same
particles at all times. At each time step the process
hosting a fluid particle gathers contributions to the interpolated
velocity, from all other processes using
a global collective communication call --- which unfortunately
may not scale well at large core counts.
A reduction in scalability at large problem sizes 
\citep{buaria2014} is thus not unexpected, which is
more problematic in studies of  
backward dispersion \citep{BSY.2015,BYS.2016} or rare 
extreme events \citep{YZS.2015}, where  a larger number of
particles are necessary.
To overcome this limitation we have devised an alternative
approach where at each time step, a particle is tracked by
the parallel process which 
holds the sub-domain where the particle is instantaneously
located. Each process is then responsible for a
dynamically evolving instead of a fixed sub-population
of particles. The communication required becomes local
in nature, occurring (if at all) only between parallel  processes holding
sub-domains adjacent to one another.
This ``local communication'' approach
has some  similarities with
the spatial decomposition techniques in 
molecular dynamics applications
\citep{Plimpton,namd}, and 
has been used for fluid and inertial particles
in turbulence simulations as well \citep{ireland2013, ayala.cpc}.
In these previous works 
usually the host process
gathers information from its neighbors in the form of
so-called ``ghost layers'' immediately outside the boundaries
of each sub-domain, in  a manner similar to
parallelized finite difference codes.
However, use of ghost layers incurs
substantial costs in memory and communication, especially
when cubic spline interpolation \citep{YP.1988} with
a stencil of $4^3=64$ points is used in conjunction 
with a domain decomposition based on Eulerian
simulation requirements at large process counts.
 
Our new algorithm reported in this paper is based on the
framework of the local communication 
approach discussed above, but avoids ghost layers completely.
Instead, we utilize
a partitioned global address space (PGAS)
programming model, namely Co-Array Fortran
\citep{Numrich98}, to fetch the required data directly
from remote memory using one-sided communication.  
In PGAS programming, the memory of all processes is treated as a global memory,
but at the same time portions of shared memory will have affinity towards particular processes,
thereby exploiting the locality of reference.
In addition,  latencies for short messages in PGAS programming 
can be significantly smaller compared to 
that provided by the standard Message Passing Interface (MPI) library.
This makes PGAS models
especially appealing in the current work, because the local communication
pattern noted above allows the code to benefit from
the memory affinity in PGAS, while the message sizes
required for interpolation are also small.

While the programming concepts involved
are general, in this work we have focused on
code performance on the petascale supercomputer
{\em Blue Waters}, which is a
Cray system consisting of 22,400 XE6 and 4,220 XK7 nodes
operated by the National Center for Supercomputing Applications (NCSA)
located at the University of Illinois at Urbana Champaign, USA.
We are able to achieve very good performance 
(with a very large speedup) for
our production problem size of $8192^3$ grid points
on 262,144 Cray XE cores (i.e. 8192 nodes)
and up to of order 300 million  fluid particles.

The rest of the paper is organized as follows.
In Sec.~2 we provide background information on the DNS code
which calculates the velocity field, and on our
baseline particle tracking algorithm based on
a global communication pattern.
We also present some performance data for this baseline
approach, showing why it is not suitable
for simulations at petascale or post-petascale problem sizes.
In Sec.~3 we discuss in detail the new parallel implementation
which is based on one-sided communication (as opposed to
ghost layers) and  local communication.
In Sec.~4 we provide a performance analysis which shows that,
somewhat counter-intuitively, scalability actually improves
as the problem size and core count increase.
Finally in Sec.~5 we summarize the performance improvements
in this work and briefly 
comment on its potential 
applicability on the next wave of
post-petascale platforms to come.
Science results enabled by the new algorithm are to be 
reported separately.

\section{Eulerian setup and base particle-tracking algorithm}

While our focus  is on tracking particles,
efficient calculation of the Eulerian velocity field
is a major prerequisite which also drives the overall structure
of the DNS code. We thus begin with a 
brief account of 
the Eulerian DNS code which calculates the velocity field.
We also give a brief account of  how cubic spline coefficients are calculated,
and of our baseline algorithm
in which the mapping between 
particles and MPI processes is fixed in time.

\subsection{Eulerian DNS code structure}

In the interest of simplifying the flow geometry but 
focused on reaching higher Reynolds numbers, we consider
stationary homogeneous isotropic turbulence in a 3D periodic domain.
The incompressible Navier-Stokes equations expressing conservation
of mass and momentum for the velocity fluctuations $\uu(\xx,t)$ can be written as
\begin{align}
\nabla \cdot \uu &= 0 \ 
\label{eq:mass} \\
\partial \uu / \partial t + \uu \cdot \nabla \uu &= -\nabla (p/\rho) +
\nu \nabla^2 \uu + \mathbf{f} \ ,
\label{eq:mom}
\end{align}
where $\rho$ is the density, $p$ is the pressure, 
$\nu$ is the kinematic viscosity
and $\mathbf{f}$ denotes a numerical forcing term used to sustain
the fluctuations \citep{EP88,DY2010}.
The solution variables are expanded in a Fourier series with 
a finite number of wavenumber modes.
The equations are transformed to wavenumber space,
 where the divergence-free condition in (\ref{eq:mass}) 
is enforced by projecting all terms transformed from (\ref{eq:mom}) onto
a plane perpendicular to the wavenumber vector. 
To avoid prohibitively-costly convolution sums associated with
Fourier transforms of the nonlinear terms 
($\uu\cdot\nabla\uu$) it is standard to use a pseudo-spectral method 
\citep{Rogallo,canuto88}, whereby the nonlinear
terms are formed in  physical space and transformed back
to wavenumber space.
The resulting aliasing errors 
are controlled by a combination of truncation and phase-shifting techniques
\citep{PO1971}.
Time integration is performed using an explicit
second order Runge-Kutta method, where a Courant 
number (C) constraint of $C < 1$  is required for  numerical
stability.

\begin{figure}[h]
\centering
\includegraphics[height=1.5in]{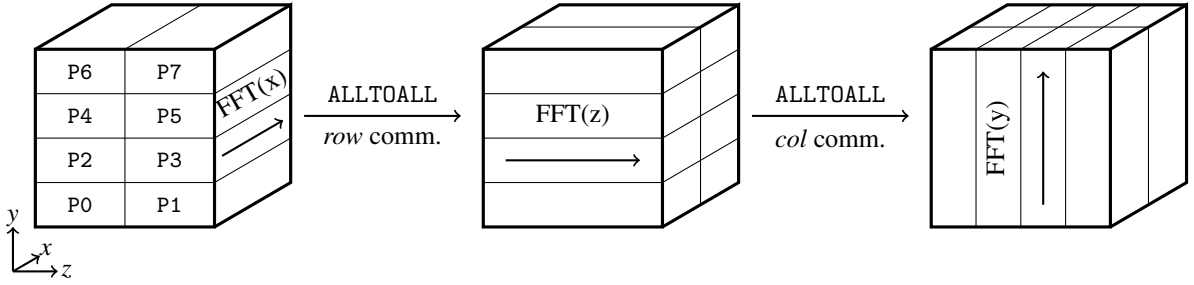}
\caption{A schematic showing the 2D domain decomposition and transposes
required for the 3D-FFT. For simplicity, we show the case of $2\times4$ processor
grid with 8 parallel processes labeled \texttt{P0} to \texttt{P7}.
}
\label{fig:pencils}
\end{figure}

Clearly, the feasibility of high-resolution 
pseudo-spectral DNS
is highly dependent on the  
parallel implementation 
of 3D FFTs, which are of general interest in themselves
\citep{DYP2008,Mininni_PC,p3dfft,Ayala_PC}. 
Our Eulerian code base is thus designed
to make the 3D FFTs as efficient as possible.
We consider the case of $N$ grid points in each direction.
The simplest domain decomposition scheme is one-dimensional (1D),
such that each process holds one  ``slab'' of data, of size
$N \times N \times N/P$, where $P$ is the number of MPI processes.
This method is obviously
restricted to $P \leq N$, which 
is less appealing on
very large systems unless OpenMP multithreading is highly effective.
Instead we use a 2D decomposition, 
such that each process holds a ``pencil'' of data, of 
size $N \times N/P_r \times N/P_c$, where $P_r \times P_c = P$
defines the 2D Cartesian process grid geometry. 
In this setup, there are $P_c$ row communicators of size $P_r$ each, and
likewise $P_r$ column communicators of size $P_c$ each
\citep{DYP2008}. 
The schematic in Fig.~\ref{fig:pencils} illustrates the sequence of
operations for  a 3D real-to-complex transform, beginning with
pencils of data in the first ($x$) direction. 
The 1D FFTs in each direction
are taken 
using the FFTW software library,
with data in the local memory, while transposes using 
\texttt{ALLTOALL} collective communication within
row or column communicators are used to re-align
pencils of data along the required directions.
Because data involved in the communication calls
reside in non-contiguous areas of memory, some local transpose
(pack and unpack) operations are also required. To facilitate fast
arithmetic we use stride-one arrays whenever possible.

Pseudo-spectral codes tend to be communication intensive.
We have found it beneficial to 
let $P_r$ be small compared to $P_c$, with $P_r$
matching (or less than) the number of cores on a node (say $P_{node}$),
such that the 
\texttt{ALLTOALL} within the row communicator
can occur entirely on the node, bypassing  the slower interconnect.
For example, on {\em Blue Waters} with $P_{node}=32$, 
we have performed $8192^3$ simulations
on a $32\times 8192$ processor grid.
We also use a PGAS implementation (based on Co-Array Fortran) 
to perform the \texttt{ALLTOALL}s \citep{fiedler.CAF}, whereby
remote-memory access (RMA) is utilized by declaring  
global co-arrays as buffers to perform the required communication.
The use of such RMA based programming, typically utilizing MPI-3 or PGAS models,
to improve collective communication costs,
is becoming increasingly widely adopted  \citep{Gerst,Nishtala}.

While the communication costs are greatly reduced by using these strategies, 
the choice of $P_r \le P_{node}$ means that $P_c$ must be made larger as $P$ increases, 
with $N/P_c$ ultimately becoming as small as unity.
As we will see later, this 
feature  has special implications for our new  particle 
tracking algorithm.

\subsection{Cubic-spline coefficients and baseline particle tracking algorithm}

Several different interpolation schemes have been used in the
literature \citep{YP.1988,HDG.2007,ireland2013,ayala.cpc} 
to obtain particle velocities in turbulence simulations.
A scheme of high order of accuracy is important, especially
if one wishes to compute velocity gradients (which are
less well-resolved in space) following the particle trajectories as well 
\citep{yeung2001,Meneveau.2011}.
In addition, the study of fluid particle acceleration
 \citep{Sawford.2003,YPKL07}, which is obtained
by differentiating the velocities in time, requires that the
interpolated functions be smooth in space.
These requirements are well met by cubic-spline
interpolation, which is fourth-order accurate 
and (in contrast to piece-wise polynomials) twice differentiable.

Suppose at any given time step, a fluid particle is located 
within a grid cell labeled by the
integer indices $\alpha,\beta,\gamma$  such that
$x_\alpha\leq x^+\leq  x_\alpha+\Delta x$,
$y_\beta\leq y^+ \leq y_\beta+\Delta y$, and
$z_\gamma\leq z^+\leq z_\gamma+\Delta z$, with
uniform grid spacings 
$\Delta x$, $\Delta y$, $\Delta z$; and $S_{pqr}$
with $1\leq p,q,r\leq N+3$ are 
the cubic spline coefficients
for a flow variable $g(\xx)$ in 3D space.
The interpolated value of $g$ at the particle position is given by
\begin{equation}
g^+ = \sum_{k=1}^4 \sum_{j=1}^4 \sum_{i=1}^4 
b_i (x') c_j (y') d_k (z') S_{pqr} \ ,
\label{eq:spline}
\end{equation}
where $p=\alpha+i-2$,
$q=\beta+j-2$,
$r=\gamma+k-2$; primes indicate  normalized local
coordinates, such as $x'=(x^+ - x_\alpha)/\Delta x$; and
$\{b_i\}$, $\{c_j\}$ and $\{d_k\}$ are 1D 
basis functions in $x$, $y$, $z$ directions respectively.
The latter are of 
compact support 
over an interval of four grid spacings only
and have the same
prescribed functional forms as in Ref.~\cite{YP.1988}.
Periodic boundary conditions 
for the particle velocities are enforced by recognizing
that the velocity of a particle located outside the primary
domain of length $L_0$ on each side (usually $L_0=2\pi$) 
is the same as if it were at a ``shadow'' location
inside the primary domain 
shifted by multiples of $L_0$ in each direction. 
For each particle, use of the interpolation formula above  requires
three major operations, which we refer to as
(1) generation of spline coefficients based on the velocity field;
(2) evaluation of basis functions based on the particle position; and
(3) summation over $4^3=64$ contributions.

The first of these three operations
is Eulerian in nature and independent
of the other two, as well as the number of particles tracked.
Similar to FFTs, the spline coefficients are partitioned
using a 2D domain decomposition and operated on one direction
at a time, in the ordering $x$, $z$, $y$ as
suggested in Fig.~1.
Along each grid line of
$N$ grid points we determine $N+3$ coefficients
by solving a tridiagonal system of simultaneous equations
\citep{Ahlberg} with periodicity in space.
Transposes between pencils of partially-formed spline
coefficients are also required. However since $N+3$
is not divisible by $P$ there is a slight imbalance in the
message sizes that each MPI process sends and receives.
Consequently the transposes
are implemented by \texttt{ALLTOALLV} constructs which allow for
non-uniform message sizes. 
On {\em Blue Waters} an improvement in performance is obtained
by using a Co-Array Fortran  equivalent of \texttt{ALLTOALLV}
\citep{fiedler.CAF}.
In addition, operations needed to solve spline equations
in the $y$ and $z$ directions
are subject to a non-unity vector stride and are hence slower than
that in the $x$ direction, while the packing and unpacking
operations are also less efficient than those used for 3D FFTs
on an $N^3$ array. For these reasons, overall the operation of
forming the spline coefficients scales less well
than 3D FFTs, and the cost of generating the spline
coefficients may be substantial. However, this cost
is a necessary expense
if differentiability of the interpolated
results is important.

Operations 2 and 3 as listed above are  dependent on how information  
on the particle population is divided among
the MPI processes. The baseline version of our algorithm
uses a static-mapping approach where (as noted in Sec.~1)
%mark 2
%each particle is tracked by the same MPI process at all times, and 
each MPI process is responsible for the same particles
at all times. The total population of $N_p$ particles
is divided into $P$ sub-populations of size $N_p/P$ each.
Initially, each
sub-population of
particles can be distributed randomly 
either within its host sub-domain, or 
throughout the solution domain, with coordinates
between $0$ and $L_0$.
The latter is convenient for post-processing,
where each sub-population of particles can be taken as a
single realization for ensemble averaging.
Statistical independence
between these sub-ensembles is achieved by  a different
random number seed for each MPI process when 
calling Fortran intrinsic random
number generator (\texttt{RANDOM\_NUMBER})
to initialize the particle positions.

Operation 2 can now be carried out readily on each 
MPI process, since the 1D basis functions 
($b_i, c_j, d_k$)
are simple algebraic functions
of the reduced particle position coordinates 
($x',y',z'$, which are already known
to the MPI process). However in preparation for
Operation 3, information on local particle coordinates
(the quantities $\alpha$, $\beta$, $\gamma$, $x'$, $y'$, $z'$ used in
(\ref{eq:spline})) for all particles must be made available
to all MPI processes.
This information sharing can, in principle, be implemented through a
global \texttt{MPI\_ALLGATHER} collective communication call,
which however scales very poorly at large $P$.
Alternatively, to reduce the number of MPI processes
engaged in collective communication we can use
a hierarchical approach based on 
a row-and-column communicator of dimensions 
say $P_1\times P_2$
(which are distinct from $P_r$ and $P_c$ used in the rest of the code).
This scheme consists of 
an \texttt{MPI\_GATHER} first used to collect 
data within each row, followed by an \texttt{MPI\_ALLGATHER}
across a column, and finally a \texttt{MPI\_BCAST} 
(which can be implemented using non-blocking
\texttt{MPI\_ISEND} and \texttt{MPI\_RECV}s)
back within each row. 
Since only one lead process from each row needs
to participate in the \texttt{MPI\_ALLGATHER}
a substantial improvement is achievable by
using a small $P_2$.
Both a standard \texttt{MPI\_ALLGATHER}
and its hierarchical version require additional storage,
which is in principle 
proportional to the number of particles but 
can be reduced by 
dividing the $N_p$ particles into several
batches and operating on each batch sequentially.

After the \texttt{MPI\_ALLGATHER} communication above, 
each MPI process is now able to participate
in Operation 3 by calculating 
its own partial contributions
to the summation in (\ref{eq:spline}) for all particles.
This task requires collecting and adding partial sums
collected from different MPI processes and subsequently
returning to each MPI process the results for the particles
that it is responsible for. 
In principle these data movements can be accomplished by 
using a combination of 
\texttt{MPI\_REDUCE} and 
\texttt{MPI\_SCATTER} between all processes.
However, we have also used CAF to implement
this \texttt{REDUCE+SCATTER} using a binary-tree communication pattern \citep{buaria2014},
whereby processes exchange information with each other in pairs over 
$\log_2P$ cycles. By using smaller message sizes, similar to CAF implementation
of \texttt{ALLTOALL} \citep{fiedler.CAF}, the one-sidedness of the tree-based CAF implementation allows 
for significant reductions in latency costs, 
thereby offering significant speedup over its MPI counterpart.

Since the operations in the two preceding paragraphs
are communication-sensitive, in the search
for improved scalability we have also considered use of
OpenMP multi-threading \citep{buaria2014}. In the  hybrid MPI-OpenMP
programming model, it is best to use a configuration such
that the product of $P_r$ and the number of threads per MPI process
($n_{thr}$) is equal to the number of cores available per node.
To avoid memory-access penalties across
different NUMA domains, $P_r$ can be set  equal to
the number of NUMA domains available on each node, with
all threads associated with a given MPI process
placed within the same NUMA domain.
%and the 
%An additional constraint can be placed on $n_{thr}$ to avoid
%memory-access penalty across NUMA domains,
%%utilize the memory affinity within a NUMA domain,
%hence restricting it to the available cores on a single
%NUMA node (and $P_r$ to the total number of available NUMA-nodes)}.
The  DNS code in our work is actually completely hybridized
for simultaneous use of MPI and OpenMP.
A reduction of the number of MPI processes does 
lead to better performance for Operations 2 and 3 
via a reduced costs in latency associated with
large process counts.
However on {\em Blue Waters} OpenMP appears to be much less
competitive when used together with Co-Array Fortran,
and the generation of spline coefficients
does not benefit from multi-threading.
Accordingly, we present only
single-threaded timings here.

As might be expected, for a given problem size, the
performance of the approach described here depends on
the number of MPI processes, the communication performance
of the machine used, as well as the availability of a robust
Co-Array Fortran implementation. 
Because of a heavy reliance on global
communication over many parallel processes, it is not
surprising that scalability is not sustained well
at large problem sizes.

\begin{table}[h]
\centering
    \begin{tabular}{l||c|c|c||c|c|c} \hline \hline
    Grid points ($N^3$)    &  $2048^3$   & $4096^3$    & $8192^3$ & $2048^3$    & $4096^3$ & $8192^3$ \\
    CPU cores ($P$)       &  4096   & 32768   & 262144  & 4096 & 32768 & 262144 \\
    Proc. Grid  ($P_r\times P_c$)  &  32x128 & 32x1024 & 32x8192 & 32x128 & 32x1024 & 32x8192 \\
    No. particles    ($N_p$)       & 16M & 16M & 16M & 64M & 64M & 64M \\
\hline 
    Eulerian (3D-FFTs)	   & 4.60 & 6.59 & 9.20 & 4.60 & 6.59 & 9.20 \\
    Weak scaling \%	   & --    & 76.1\%& 77.6\%& --    & 76.1\% & 77.6\% \\
\hline 
    Spline coefficients    &  1.66  & 2.33   & 4.42  & 1.66 & 2.33  & 4.42  \\
    Weak scaling \%        &  --     & 71.2\%  & 52.7\% & --    & 71.2\% & 52.7\% \\
\hline 
    Allgather (global)     &  0.71  & 2.89   & 7.69   & 2.83 & 11.32 & 29.57  \\
    Allgather (hierarchical) &  0.70  & 1.03   & 1.35  & 2.99 & 4.73  & 5.58 \\
    Computations           &  0.40  & 0.39   & 0.39   & 1.68  & 1.59 & 1.60 \\
    Reduce+Scatter (CAF)   &  0.83  & 1.52   & 2.94   & 4.14 & 8.32 & 11.34 \\
    Particles total        & 1.93 & 2.94 & 4.68 & 8.65 & 14.64 & 18.52 \\
\hline 
    Interpolation total    &  3.59  & 5.27  & 9.10 & 10.31  & 16.97 & 22.94  \\
\hline
\hline
    \end{tabular}
\caption{Performance data obtained on {\em Blue Waters}
using the static particle-to-process mapping  in our baseline algorithm 
for $N_p$=16M (left columns) and  64M (right columns) particles (where $M=2^{20}=1,048,576$).
The timings shown are elapsed wall time per second-order Runge Kutta
time step, to the nearest hundredth of a second. Weak scaling is calculated
for each doubling of $N$,
with the number of operations
being proportional to $N^3\log_2 N$ for the Eulerian code and
$N^3$ for the calculation of spline coefficients.
The ``particles total'' entry is the sum of Allgather (hierarchical),
computations and reduce+scatter (using CAF).
The ``Interpolation total'' entry is the sum of ``particles total'' and
spline coefficients.
}
\label{tab:global}
\end{table}

Table~\ref{tab:global} gives a brief summary of performance data
for the baseline algorithm. The data are collected by, as usual,
measuring the time elapsed between suitably-placed
\texttt{MPI\_WTIME} calls, for the slowest MPI process but
taking the best timing
over a substantial number of time steps or iterations.
For each value of $N_p$, the Eulerian problem size
is varied from $2048^3$ to $8192^3$, while the number of 
MPI processes ($P$) is varied in proportion to $N^3$.
We also report weak scaling over each doubling of $N$,
considering the differences in  operation counts
for 3D FFTs and for the calculation of spline coefficients
(based on the solution of tridiagonal systems
using the well-known Thomas algorithm). 
The last row of the table is the sum of all contributions to
the cost of interpolation, including generation of spline
coefficients, hierarchical allgather, computation, and the 
tree-based Co-Array Fortran implementation of 
\texttt{REDUCE+SCATTER}
for assembling results for all particles and 
re-distributing them back among
the MPI processes.

The Eulerian parts of the code have been optimized aggressively
in support of recently published work that
did not involve fluid particles \citep{YZS.2015}. About 77\% weak scaling is obtained
for each doubling of $N$. As suggested
earlier in this subsection the calculation of spline coefficients 
scales less well while being also independent
of $N_p$.
In subsequent rows of the table the timings are
generally proportional to $N_p$.
A global \texttt{MPI\_ALLGATHER}
over all MPI processes is seen to perform very poorly, with
roughly a factor of 3 increase in cost between successive problem sizes.
The hierarchical scheme performs much better but its scalability
is also not good, considering that it takes 
longer even as the number of
particles per MPI process decreases by a factor of 8
between adjacent columns of the table with $N_p$ held fixed.
Scalability measures of the computational operations in (\ref{eq:spline})
and subsequent communication needed to
complete the interpolation
are also evidently far from ideal.

Since our science objectives call for a simulation with $N=8192$
and (at least) $N_p=256$M, four times more than the largest $N_p$
shown in Table~\ref{tab:global}, it is clear that a new
approach for the particle tracking algorithm is required.

\section{Dynamic particle-to-process mapping and local communication}

As indicated earlier (Sec.~1), to reduce communication costs in interpolation
it is helpful to divide the particle population among the MPI processes
according to a dynamic mapping based on instantaneous particle positions, such
that at each time step each particle is processed by the MPI process
that holds the sub-domain where the particle is located.
If the interpolation stencil (of 64 points for 
cubic splines) surrounding the particle position
lies wholly within the sub-domain then no communication is needed.
Otherwise, for particles located close to the sub-domain
boundaries, communication is still required to access 
some of the spline coefficients held by one or more  neighboring MPI processes.
However, in contrast to the global communication pattern
in the baseline algorithm (Sec 2.2) these communications
will now be local, occurring only between pairs of
MPI processes holding sub-domains next to each other.
In principle, information from neighboring MPI processes can
be obtained through the concept of ghost layers, which is
common in parallelized finite difference schemes but 
(as explained below) is not 
ideal for our application.
Instead we have devised a new communication protocol based
on one-sided communication via Co-Array Fortran (CAF), which
is advantageous on {\em Blue Waters} and likely to be
more widely available in the future.

An inherent feature of the dynamic particle-to-process mapping is that,
as particles cross the sub-domain boundaries, control for
the migrating particles needs to be passed from one host MPI
process to another.
Since the number of particles tracked by each MPI process
now changes dynamically at every time step, some transient load
imbalance is anticipated. However, since in
homogeneous turbulence the spatial distribution of particles
is statistically uniform, this imbalance is expected to be minor,
as long as the average number of particles per MPI process
is large ($N_p/P=1024$ in our largest simulation).
At the same time, only particles already located very
close to the sub-domain boundaries can possibly migrate,
and the likelihood of such migrations is proportional 
to the time step size ($\Delta t$). 
In our simulations, for reasons of numerical stability and temporal
accuracy, we choose $\Delta t$ such that
the Courant number to be 0.6 --- which
means no fluid particle can travel more than 0.6  grid spacing
in any coordinate direction
over one time step. This constraint is expected to 
help reduce  the communication overhead
for the inter-process migrations.

The principle of ghost layers is that each parallel process
extends its reach by accessing several layers of information along the
boundary with a neighboring  parallel process, while also providing 
similar information to the latter.
The communication is generally
of the \texttt{SEND}+\texttt{RECV}, or halo type. Spline coefficients in
these ghost layers will have to be refreshed
--- via communication --- at every Runge-Kutta
sub-step. Since spline coefficients are stored in a 2D
domain decomposition, each MPI process will be performing these halo
exchanges with four of its neighbors (two in each direction
where the domain is sub-divided). For cubic splines these
ghost layers must  also
be three points deep on each side.
For an $N^3$ problem with $(N+3)^3$ spline
coefficients on a $P_r \times P_c$ processor grid,
since $N+3$ is not divisible by $P_r$ or $P_c$,
each parallel process needs to hold up to
$(N+3)(N/P_r+1)(N/P_c+1)$ of the coefficients. 
If ghost layers are included 
this increases
to $(N+3)(N/P_r+6)(N/P_c+6)$  based on the considerations above.
The consequent increase in memory requirements 
may be mild if both $N/P_r$ and $N/P_c$
are large numbers  (being least if 
$P_r \approx P_c \approx \sqrt{P}$),
but very substantial if one of them is small.
However at large problem sizes in the DNS,
FFT and Eulerian
code performance favors processor grids
where $P_c\gg P_r$. 
In our $8192^3$ simulation $P_r=32$ and $P_c=8192$,
such that $N/P_c$ is as small as unity. This implies
a memory increase by a factor of 
at least $(1+6)/(1+1) = 3.5$, which will likely require
using even more cores and thus make the simulations
more expensive.

In addition to memory,
the size of the ghost layers
also has a direct effect on the volume and cost of
the communication traffic involved
in the halo exchanges. However, each spline coefficient
in those ghost layers will  be actually used for interpolation
only if there is at least one particle in the pertinent immediate
neighborhood. The probability of such an 
occurrence is proportional to the number of particles
per grid point, i.e. the ratio $N_p/N^3$,
and the fraction of particles located within three  grid spacings
of the sub-domain boundaries.
Although in general the number of particles
needed for reliable sampling increases with Reynolds number 
\citep{yeung2002,BSY.2015},
in most simulations
$N_p$ is much smaller than $N^3$. Indeed, in  
our $8192^3$ simulation even with $N_p$ in the order of
$3\times 10^8$,  the ratio $N_p/N^3$ is still less than 0.001.
Furthermore, with $N/P_r$ being large,
the fraction of particles that actually require spline coefficients
in the ghost layers extending in the direction of the row communicator
is expected to be small.
Consequently, in contrast to finite difference calculations, for our
application both the memory and communication costs of having
the complete ghost layers are very wasteful. 
The effects of additional memory may be less severe if we use a 3D domain
decomposition, since then no individual dimension of the 3D sub-domains
will be particularly small. However this will prevent FFTs from being taken
in core, and hence would adversely affect the performance of the Eulerian portions
of the code.

To scale effectively to extreme problem sizes, we have 
designed a new algorithm that avoids ghost layer completely,
while associating each particle dynamically with the MPI process
that holds the sub-domain where the particle resides.
To avoid the wastefulness
of unused spline coefficients in the ghost layers we
transfer  spline coefficients between
neighboring parallel processes only on an as-needed basis.
This is achieved by examining the position of each particle, and
deciding (based on the proximity of the particle to 
sub-domain boundaries), 
whether any (and how many) spline coefficients are needed from neighboring
MPI processes. The host MPI process for the particle then
fetches only those specific coefficients,
directly from remote memory using one-sided communication. 
After this task the host process is now able to
complete the interpolation for the particle velocity
and calculate updated position coordinates. If a particle
has migrated to an adjacent sub-domain 
a one-on-one halo exchange  is  also used to 
transfer the control of the particle to its new
host MPI process.

The key to the performance of this new algorithm is obviously
in how the one-sided communication is performed. 
For each particle, the 64 spline coefficients needed are either
available in the local memory or fetched from the remote memory
and stored in a temporary array.  
To fetch the spline coefficients efficiently,
we use a partitioned-global address space (PGAS) programming model,
such as Co-Array Fortran (CAF). 
Essentially, the entire 
memory space of all the processes is treated as a global memory,
partitioned logically such that a portion of it is local
to each MPI process. However, 
(as noted in Sec.~1) depending on the physical proximity
to the memory of each process, portions of the shared memory
space may have an affinity for a specific parallel process.
This suggests the memory locality of the data can be exploited
for further optimization.
In CAF, which is well supported on {\em Blue Waters}, 
the PGAS model is implemented by declaring
global co-arrays, which have an additional co-dimension
(denoted by square brackets, 
distinct from the usual parentheses
for regular arrays), which allows any MPI process  to access information
held by other processes. 
Compared to MPI, communication calls
in CAF can (due to one-sidedness) 
%can %mark 2 
have smaller headers 
and therefore can carry more
data per packet for slightly higher bandwidth.  
Latencies for short messages
in CAF are also significantly lower than in MPI.
Thus CAF 
is perfectly suited for current application, 
since the messages for individual particles are small.
The performance improvement from  PGAS programming is well known
for many applications ranging from tuning of collective communication calls
\citep{fiedler.CAF, Nishtala} to molecular dynamics simulations 
\citep{Teijeiro}.

\begin{figure}[!ht]
\centering
\includegraphics[height=5.2in]{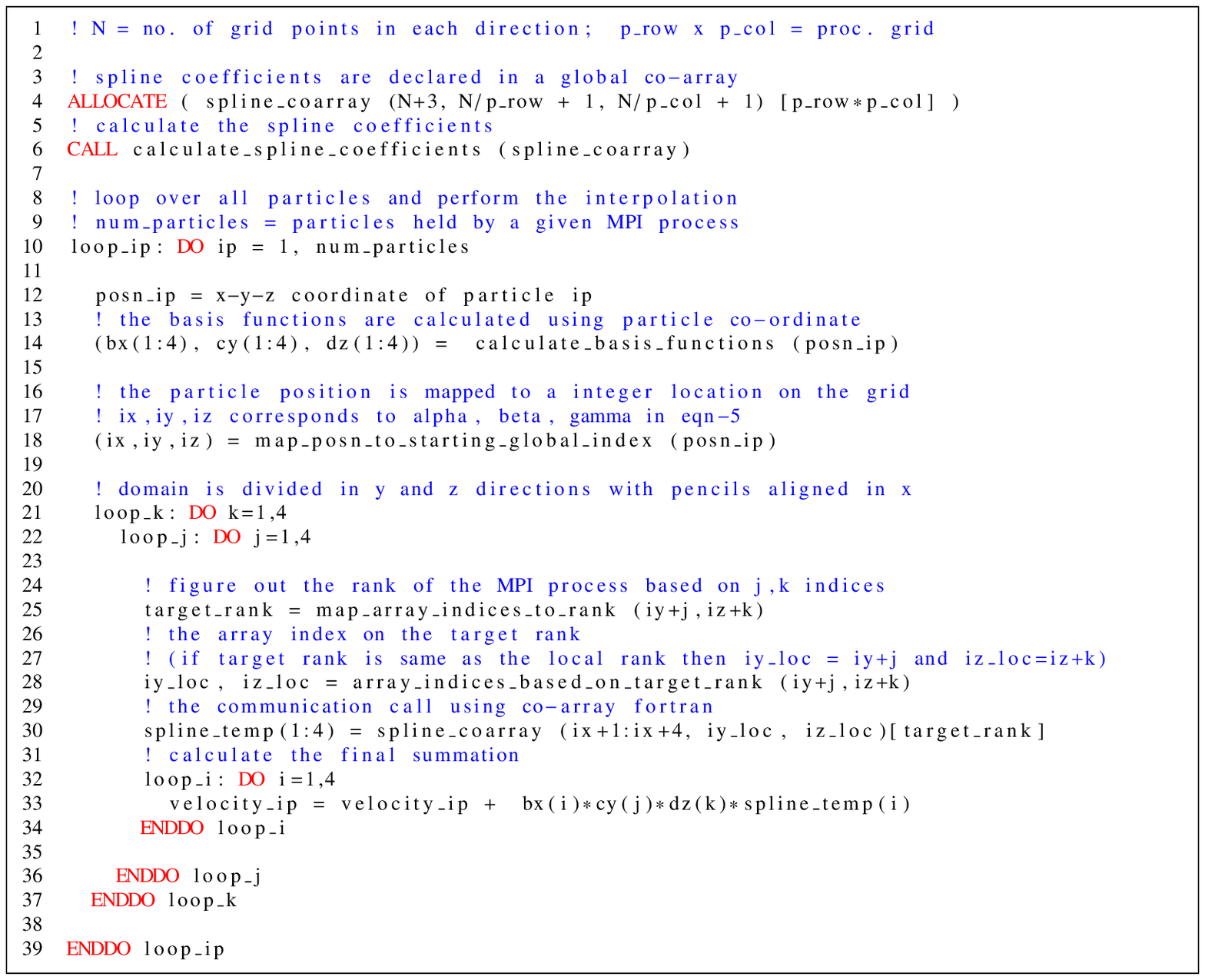}
\vspace{0.5mm}
\caption{Pseudo-code showing an outline of the sequence of
operations in cubic spline interpolation based on a spatial
decomposition of particles and local communication 
implemented using Co-Array Fortran. Fortran syntactical
elements are in red; descriptive comments are in blue.}
\label{fig:code}
\end{figure}

Our strategy of using CAF for the interpolation is illustrated
by a pseudo-code shown in figure~\ref{fig:code}. 
We begin at
a stage in the calculation where the Eulerian velocity field
at $N^3$ grid points is available as pencils
sub-divided in the $y$ and $z$ directions.
The array containing the spline coefficients
is declared as a global array, called \texttt{spline\_coarray}.
The last dimension of this array,
in square brackets, is the co-dimension, which is the same
as the rank (0 to $P-1$) of each of the
$P$ MPI processes used. As noted in Sec.~2.2, the spline 
coefficients are calculated by solving a tridiagonal system of equations
in each direction, with a total of two \texttt{ALLTOALLV}
transposes in between, one for each of the row and column 
communicators.
The code then enters the main interpolation section which
loops over all the particles.
For each particle we map its position coordinates
to  indices $\alpha,\beta,\gamma$ and the normalized 
local coordinates $x',y',z'$ that together allows
the basis functions required in (\ref{eq:spline}) to be calculated.
This mapping is direct and simple, but also accounts
for particle positions outside  the primary domain 
of dimensions $L_0^3$, through a modulo function which
gives a corresponding position inside the primary domain by
adding or subtracting a multiple of $L_0$
in each direction.

The most important use of CAF in our algorithm is to fetch
the spline coefficients required by each particle.
For each particle the code loops over the four
(with the indices $j$ and $k$ from 1 to 4)
basis functions in each of the $y$ and $z$ directions.
For each choice of $j$ and $k$, the code calculates the rank
(denoted by \texttt{target\_rank}) of the MPI process
holding the spline coefficients required.
If a particle is lying in the interior (at least
four grid spacings from the boundary) of the sub-domain
of its host MPI process then \texttt{target\_rank}
would be the same as the rank of the host process. Otherwise,
it will be the rank of one of the neighboring processes, from which
some spline coefficients are to be fetched,
along with corresponding local indices.
In each case the spline coefficients are to be transferred
in packets of four, covering all information
needed in the $x$ direction (which is
not sub-divided). 
The message size in each 
CAF memory copy operation is thus
very small (4 floating-point words), 
for which CAF usually performs best.
The co-array language syntax
is very convenient, in that
if the \texttt{target\_rank} as shown in figure~\ref{fig:code}
is the same as the rank of the local MPI process,
then the CAF assignment operation
functions as a local memory copy; otherwise, 
a one-sided communication
is performed to fetch the required information from the 
global memory.
The \texttt{target\_rank} process  is always a neighbor of the
host MPI process, which
allows us to further exploit the memory locality of the
co-array, thus achieving fast and efficient communication.
After this communication is complete the summation
over 64 basis functions and spline coefficients
in (\ref{eq:spline}) is performed entirely 
by the host process for each particle.

Before proceeding to the performance results in 
the next section (which shows very favorable
performance for our CAF-based approach), it is worth 
noting that there may be some applications where the ghost
layer approach may prevail instead. The ghost layer
approach is bandwidth-bound, with communication cost proportional
to the number of grid points ($N^3$); while the CAF approach
is latency-bound, with cost proportional to the number
of particles ($N_p$). If the particle density is very high
i.e. $N_p \gtrsim N^3$, then the CAF approach
would be inefficient since it will likely be fetching the
same spline coefficients multiple times without recognzing
that they can be fetched just once and re-used.
However in most turbulence simulations (including ours), the motivation 
for large $N_p$ comes from statistical sampling, and 
a very high particle density would imply many particles being initially
close together and thus not acting as independent samples.
As a result, even although a larger $N_p$ is desired
at large $N$, the ratio $N_p/N^3$ is very small; in fact
smaller at larger problem sizes (in our largest production science
work at $8192^3$ resolution with 256~M particles the ratio
is as small as 1/2048). 
Consequently, we anticipate our CAF would be
well suited for traditional particle-tracking turbulence simulations.
and less so if a very high particle density is
required for other science reasons (such as, perhaps,
the study of sandstorms with inertial particles).

\section{Performance and scalability analysis}

In this Section we present performance and scalability data  for our
latest particle tracking algorithm. During our recent work,
the {\em Blue Waters} machine noted earlier
was the only platform  of sufficient 
capacity %available %mark2 
available to us 
to support production simulations at $8192^3$  resolution.
Co-Array Fortran is at present particularly well supported
in the Cray Compiling Environment. For these reasons
we discuss here performance on {\em Blue Waters} only,
while also hoping to provide a reference point
for other comparisons in the future.

Since our application is entirely CPU based, we use only the
XE6 nodes which provide 32 cores per node.
The compute nodes are interconnected with a 3D torus network
topology using the Cray-Gemini interconnect.
The communication performance of the code on {\em Blue Waters} also
benefits substantially from availability  of
Topologically Aware Scheduling, which attempts to assign
to a user's job a set of nodes 
with more favorable
network topology less prone to network contention
from other jobs running concurrently on the system.
\citep{fiedler.CAF}. 
Clearly, the time taken at each time step depends on 
the number of grid points ($N^3$), the number
of particles ($N_p$), %mark2
the number of 
MPI processes ($P$), and the shape of the
processor grid ($P_r \times P_c$) used for
2D domain decomposition.
While certain parts of the particle tracking algorithm
can be timed via a separate kernel,
the actual communication traffic in the algorithm of
Sec.~3 is to some degree sensitive to the 
time evolution of the flow physics itself.
We thus report per-step timings
directly from the production DNS code, by averaging
over a large number of time steps.  This averaging also
indirectly absorbs the long-term effects of variability
due to random factors such as network contention.

In the subsections below we consider separately code
performance for generating cubic spline coefficients, using
them to obtain interpolated particle velocities,
managing the inter-process transfer for particles
migrating between adjacent sub-domains; and, ultimately, the
total simulation time per time step up to 256 M particles
on a $8192^3$ grid.

\subsection{Calculation of spline coefficients}

The calculation of $(N+3)^3$ spline
coefficients from the velocity field known at $N^3$ grid
points shares some similarities but also some significant
differences with the Eulerian 3D FFT operations.
The similarities include operating one direction
and a time, using a 2D domain decomposition, and the need for
transposes along each of the sub-divided directions.
The optimal shape of the 2D processor grid is also likely
to be the same as that for the 3D FFTs. Along each direction,
instead of FFTW we solve a tridiagonal system of equations
whose operation count scales with $N$.
The dominant cost is communication, while local
packing and unpacking also takes significant time.
However, because $N+3$ is not divisible by $P_r$ nor $P_c$
the data structure 
%is %mark2 
for spline coefficients is more intricate.
In contrast to the FFT routines the arithmetic here is
performed with unit stride only in the $x$ direction,
while operations along $y$ and $z$
have vector strides
proportional to $N$ and $N^2/P_r$ respectively.

\begin{table}[h]
    \begin{tabular}{c||ccc|ccc|cc}
    \hline
    \hline
    $N$      & 2048        & 2048         & 2048          & 4096         & 4096          & 4096          & 8192          & 8192          \\
    $P$      & 2K          & 4K           & 8K            & 16K          & 32K           & 64K           & 128K          & 256K          \\
    $P_r \times P_c$ & $32 \times 64$ & $32 \times 128$ & $32 \times 256$  & $32 \times 512$ & $32 \times 1024$ & $32 \times 2048$ & $32 \times 4096$ & $32 \times 8192$ \\ \hline
    $x$      & 0.152       & 0.071        & 0.036         & 0.152        & 0.077         & 0.035         & 0.153         & 0.076         \\
    $y$      & 0.290       & 0.149        & 0.075         & 0.309        & 0.149         & 0.075         & 0.321         & 0.160         \\
    $z$      & 0.327       & 0.160        & 0.081         & 0.520        & 0.251         & 0.139         & 0.894         & 0.453         \\
pack+unpack  & 0.281       & 0.143        & 0.076         & 0.281        & 0.144         & 0.070         & 0.290         & 0.149         \\
 alltoallv1  & 0.560       & 0.365        & 0.206         & 0.562        & 0.374         & 0.144         & 0.611         & 0.381         \\
 alltoallv2  & 1.349       & 0.761        & 0.496         & 1.976        & 1.303         & 0.828         & 4.181         & 3.215         \\ \hline
    total  & 2.990       & 1.664        & 0.977         & 3.837        & 2.332         & 1.310         & 6.449         & 4.422         \\ \hline
    \%comm. & 63.8\%      & 67.7\%       & 71.9\%        & 66.1\%       & 71.9\%        & 74.2\%        & 74.3\%        & 81.3\%        \\
    strong & --          & 89.9\%       & 76.5\%        & --           & 82.2\%        & 73.2\%        & --            & 72.9\%        \\
    weak   & --          & --           & --            & 77.9\%       & 71.3\%        & 74.5\%        & 46.4\%        & 37.6\%        \\ 
\hline
\hline
    \end{tabular}
\caption{Elapsed wall time for the calculation of spline coefficients
including a breakdown into several sub-contributions as discussed in the text.
In the leftmost column 
$x,y,z$  represent time for calculating 1D spline coefficients
in the respective directions. 
Communication costs for transposes in the row and column communicators
are in rows labeled by
\textit{alltoallv1} and \textit{alltoallv2} respectively.
(Note K denotes $2^{10}=1024$.)
}
\label{tab:spline}
\end{table}

Table~\ref{tab:spline} shows the costs of various sub-operations
in the calculation of the spline coefficients, for
problem sizes $2048^3$, $4096^3$ and $8192^3$.
For each choice of $N$ the core count $P$ is
varied over a factor of up to 4, with $P_r$ fixed at 32
(the number of cores available on each
{\em Blue Waters} node).
Because of differences in the striding,
operations in the $x$ direction
are fastest, followed with those in $y$ and $z$,
especially at
larger  problem sizes.
As the core count $P$ increases,
essentially perfect strong scaling is observed for 
these 1D operations
as well as the memory copies (packing and unpacking).
The most expensive operations are the transposes (coded as \texttt{ALLTOALLVs}),
The first transpose is faster
since it takes place in a smaller communicator
on the node, whereas the second is slower since it has
to be routed through the network
interconnect.
The percentage of time spent in communication also
increases with both problem size and core count, leading to
a gradual reduction of scalability.

\begin{figure}
\centering
\includegraphics[height=2.5in]{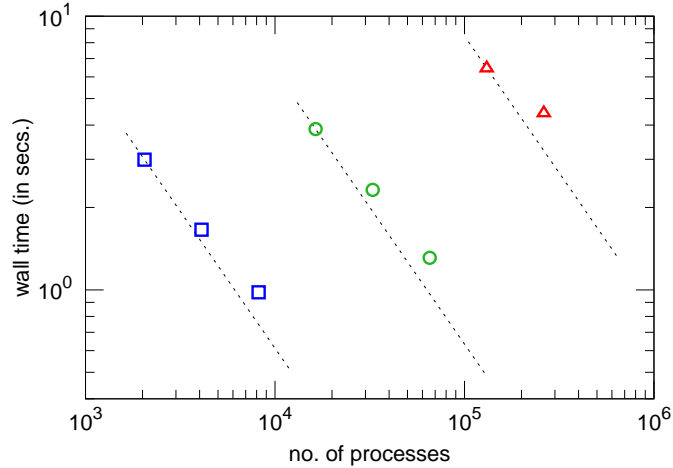}
\vspace{0.5mm}
\caption{
Elapsed wall time for calculation of
spline coefficients, versus core count ($P$) for
problem sizes
$2048^3$ (\protect\osqr{blue}),% \textcolor{blue}{$\boldsymbol\square$}), 
$4096^3$ (\protect\ocir{mygreen}), %\textcolor{mygreen}{$\ocircle$}), 
$8192^3$ (\protect\ottru{red}). %\textcolor{red}{$\triangle$}). 
Dashed lines of slope -1 on
the logarithmic scales represent ideal strong scaling with 
respect to case of smallest $P$ for each problem size.
}
\label{fig:spline}
\end{figure}

Figure ~\ref{fig:spline} shows the timings versus 
core count, on logarithmic scales where
perfect strong scaling  would be indicated by 
a line of slope -1, while
perfect weak scaling would be 
indicated by wall time being constant if $P$ is varied
in proportional to $N^3$. 
In general for a given problem size, as core count
is increased, the data points increasingly deviate from the ideal
strong scaling.
The percentages of strong scalability
shown are slightly lower than that  usually achieved
for 3D FFTs under similar conditions. The departure from
perfect weak scaling is evidently more pronounced,
especially at $8192^3$.
It is possible that aggressive use of OpenMP 
multithreading with dedicated threads for communication
\cite{ClayCPC}
may lead to some improvements in the future.
However, as shown later in the paper the overall scalability
of our new parallel algorithm is still good.

\subsection{Interpolation operations for particles}

With spline coefficients obtained as above, the cost of the
remainder of the interpolation operations is expected to
scale with the number of particles ($N_p$).
However, actual timings (with $N_p$ fixed) still
show sensitivity to the Eulerian problem size  and its
associated 2D domain decomposition.
These effects are felt through the cost of memory access
to larger arrays of spline coefficients, and --- more importantly ---
the need for communication between adjacent MPI processes
for particles lying close to the sub-domain boundaries.
The last of these effects is the most subtle and requires
careful discussion, as given in a later part of this subsection.

\begin{figure}[h]
\centering
\includegraphics[height=2.5in]{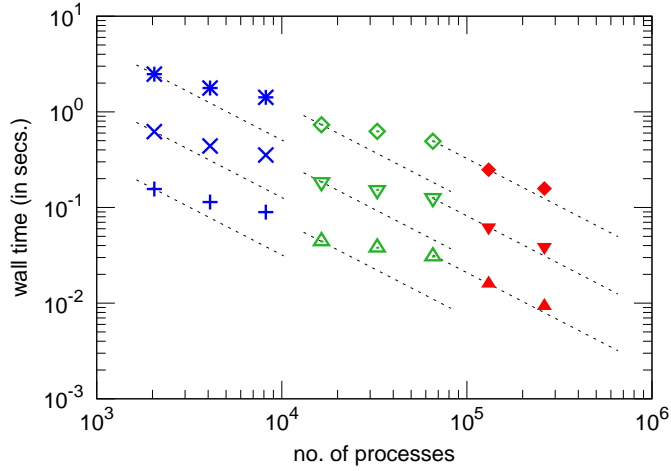}
\vspace{0.5mm}
\caption{
Elapsed wall time for the interpolation operation using the dynamic particle-to-process
mapping at grid sizes of
$2048^3$ (\protect\pluss{blue}, \protect\cross{blue}, \protect\starr{blue}),
$4096^3$ (\protect\ottru{mygreen}, \protect\ottrd{mygreen}, \protect\orhom{mygreen}), 
$8192^3$ (\protect\fttru{red}, \protect\fttrd{red}, \protect\frhom{red}),
where the three symbols for each grid size corresponds to
16M, 64M and 256M particles respectively (M = $1,048,576$). 
Dashed lines of slope -1 represents ideal strong scaling with 
respect to first case of each problem size.
}
\label{fig:interp.time}
\end{figure}

For an overview of interpolation performance we 
show in Fig.~\ref{fig:interp.time} the elapsed wall time
taken by the loop \textit{loop\_ip} in the
pseudo code presented earlier in Fig.~\ref{fig:code}. 
Since the efficiency of Eulerian operations is still important,
we consider the same grid resolutions and core counts (along with the
processor grid) as those discussed
earlier in Sec.~4.1. For each combination of $N$ and $P_r \times P_c=P$
we have obtained timings for 
$N_p=$16M, 64M and 256M,
represented by symbols
of different shapes for each color.
It can be seen that, with $N$ fixed (considering
symbols of a given color), strong scaling with
respect to $N_p$ as $P$ increases is less than perfect.
However this strong scaling improves with grid resolution:
e.g. with $N_p$ fixed the
timing for $N=4096$ and $P=16384$ is almost exactly half of that for
$N=2048$ and $P=8192$.
More significantly, the best strong scaling
with respect to $P$ 
(with both $N$ and $N_p$ fixed) occurs at
the largest $N$, which can be seen in 
the positions of symbols in red relative
to dashed lines of slope -1 in the figure.
These trends are seen to hold uniformly for all
three values of $N_p$ tested, with all timings
in the figure being proportional to $N_p$.
We explain these trends below by analyzing the effect
of the grid resolution and processor grid geometry
when employing the dynamic particle-to-process
mapping described in Sec.~3.

The elapsed wall times shown above can be separated into
contributions from computation and communication,
which we denote by $t_{comp}$ and $t_{comm}$ respectively.
The computation carried out by each MPI process is to
determine the local coordinates and perform the summation
in (\ref{eq:spline}) for each particle that is held by
that process at the beginning of a time step. This implies
$t_{comp}$ is, in principle,  proportional to the number of particles
per process, i.e. $N_p/P$, or at least will increase systematically
with $N_p$ but decreases with $P$. Because the turbulence in our
simulations is homogeneous, the distribution of particles 
%in space
is spatially uniform. Consequently, although particle migrations
between adjacent sub-domains occur on a regular basis, the
number of particles carried by each MPI process is 
expected to deviate only slightly from the averaged value ($N_p/P$).

The communication time $t_{comm}$ is, in contrast, directly related to
the number of particles for which communication is required
to  access at least some of the spline coefficients.
Referring back to the pseudo-code in Fig.~\ref{fig:code}
the key is how many times that the \texttt{target\_rank}
is not the same as the rank  of the host MPI process.
The likelihood of such occurrences is in turn tied to
the fraction of (on average) $N_p/P$ particles per MPI process
that happen to be located within some $\Gamma$ grid
spacings from the sub-domain boundaries.
(If a particle lies outside the domain itself we use
a corresponding ``shadow'' position that is inside
the solution domain and hence inside one of the sub-domains.)
Since cubic spline interpolation requires 4 points in each
direction we may take $\Gamma=2$ (half of 4).
In our 2D domain decomposition, each sub-domain is
nominally a pencil:  a cuboid measuring 
$N$, $N/P_r$ and $N/P_c$ grid spacings in the
$x,y,z$ directions respectively. Each sub-domain
has four rectangular faces in contact
with its neighbors:  two each of dimensions
$N\times N/P_r$ and 
$N\times N/P_c$ respectively.
The volume of the region where particle positions
can lead to a requirement for communication is thus
equal to $2 \Gamma (N \times N/P_r \ + \  N\times N/P_c)$
grid cells.
Since in homogeneous turbulence the particles are uniformly
distributed in space we can estimate 
\begin{align}
t_{comm} \ \propto \
2\Gamma~\frac{N \times \left( N/P_r + N/P_c \right)}{N \times N/P_r \times N/P_c} 
\frac{N_p}{P}   \ .
\label{eq:tcomm2}
\end{align}
Noting that $\Gamma$ is a constant, we can also
reduce this formula to
\begin{align}
t_{comm} \ \propto \
\frac{N_p}{P} \ \frac{P_r + P_c}{N} 
\ ,\ \  {\rm or} \ \ \ 
t_{comm} \ \propto \
\frac{N_p}{N} \ \frac{P_r + P_c}{P} \ . 
\label{eq:tcomm_a}
\end{align}
Although useful, this formula assumes
both $N/P_r$ and
$N/P_c$ are greater than $\Gamma$, which 
may not hold at the largest problem sizes.
In particular, in our $8192^3$ simulations on a 
$32\times 8192$ processor grid $N/P_c=1$, which means
each pencil is only one grid spacing wide in the $z$ direction.
Communication is then always necessary --- at least to collect the required
spline coefficients in the $z$ direction for every  particle.
In that case, one can simply write
\begin{align}
t_{comm} \propto  N_p/P \ ,
\label{eq:tcomm_b}
\end{align}
which incidentally implies perfect strong scaling,
as for the computational cost $t_{comp}$.

\begin{figure}[h]
\centering
\includegraphics[height=2.in]{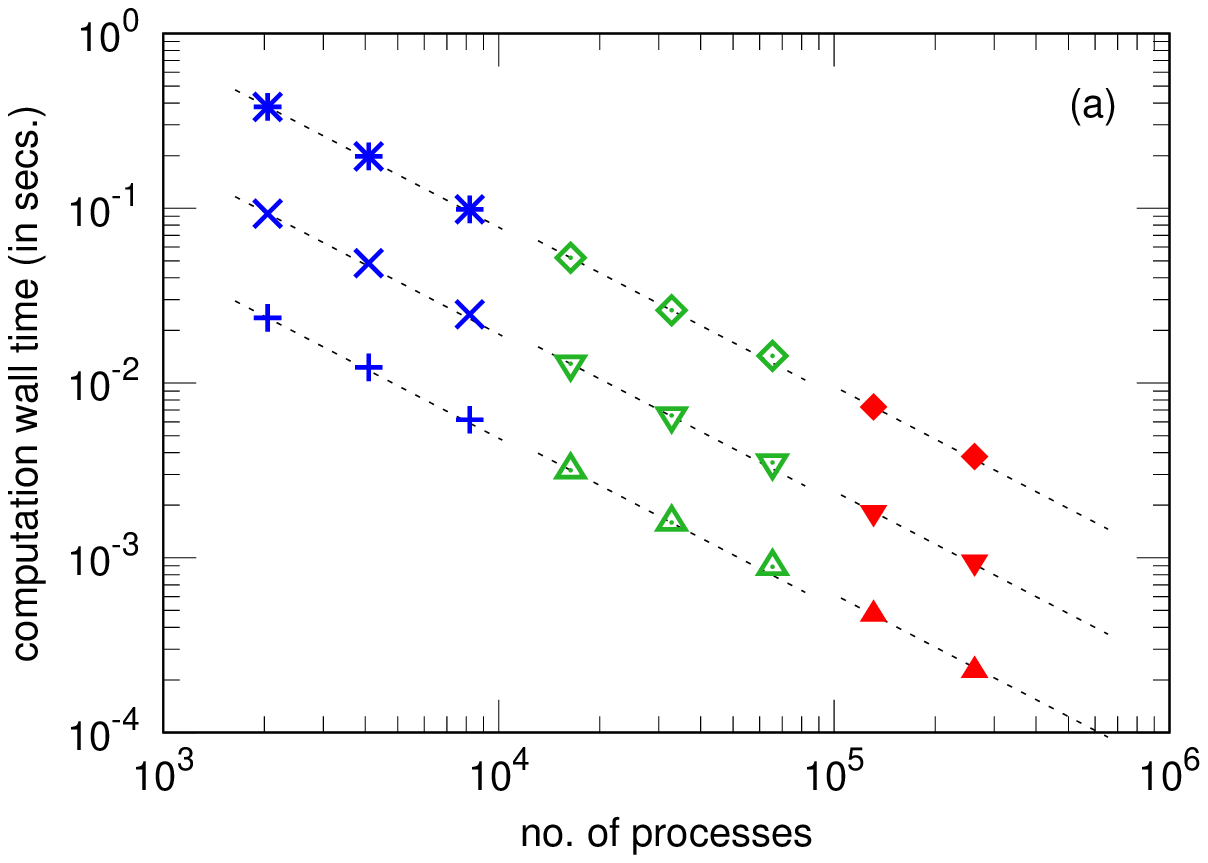}
\includegraphics[height=2.in]{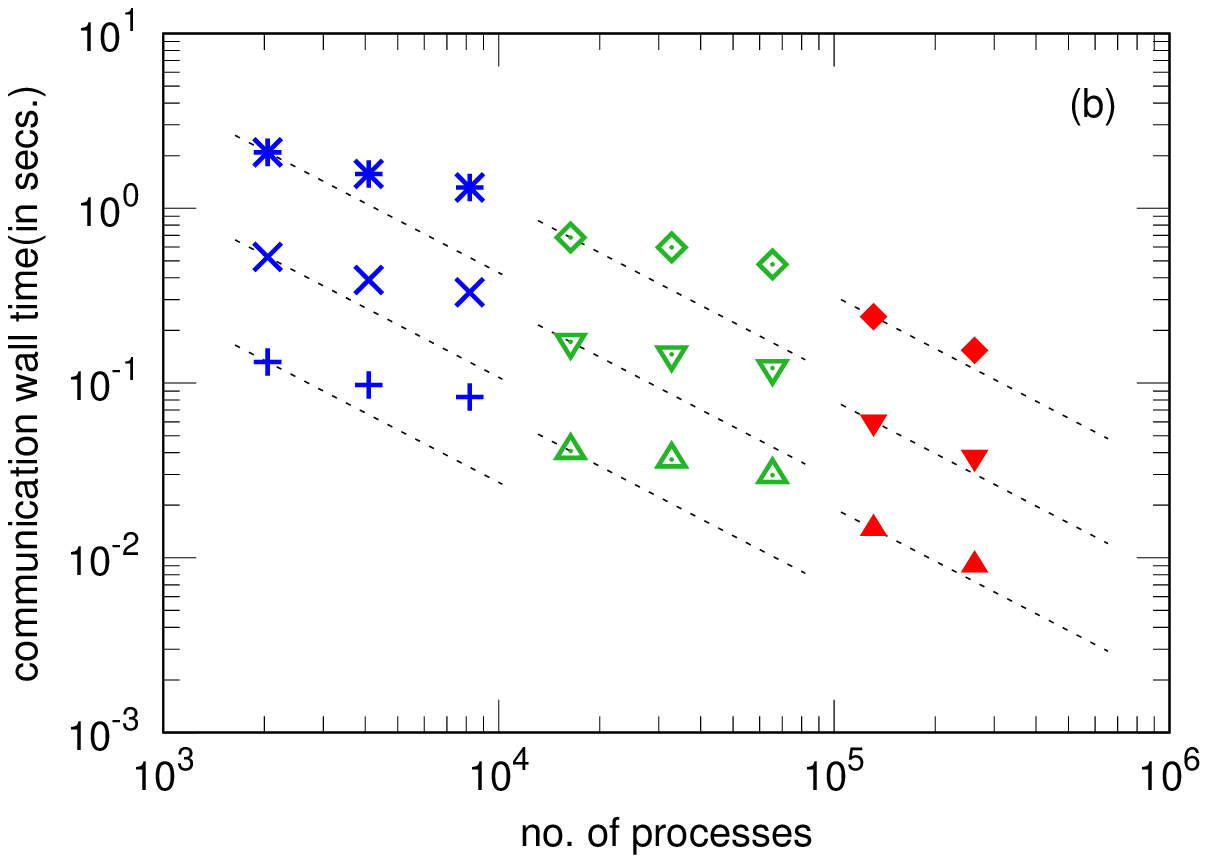}
\vspace{0.5mm}
\caption{
Breakup of (a) computational and (b) communication
cost of the interpolation. Same symbols as
Figure~\ref{fig:interp.time}.
Dashed lines of slope -1 represents ideal strong scaling with 
respect to first case of each problem size.
}
\label{fig:break}
\end{figure}

Figure~\ref{fig:break} shows a scalability plot similar
to that seen earlier in Fig.~\ref{fig:interp.time} but now
for (a) computational and (b) communication times
$t_{comp}$ and $t_{comm}$ separately.
It is clear that the computation scales almost perfectly,
i.e. $t_{comp}\propto N_p/P$ such that all data points
with the same $N_p$ lie virtually on the same line of
slope -1, with the grid resolution having only a minor effect.
Computational times for configurations with the same $N_p/P$,  
i.e. same nominal number of particles per process are also
seen to agree closely. A mild increase in such timings
at higher $N$ (e.g. the dashed line for 
$8192^3$ data points is slightly shifted upwards) 
is likely to be the result of increasing vector strides in 
arrays holding the spline coefficients accessed by the code.
Some minor variations due to deviations from
perfect load balancing caused by particle migrations
are also expected. This latter effect is also more significant
when $N_p/P$ is reduced.

\begin{table}[h]
    \begin{tabular}{c||ccc|ccc|cc}
    \hline
    \hline
    $N$     & 2048        & 2048         & 2048          & 4096         & 4096          & 4096          & 8192          & 8192          \\
    $P$      & 2K          & 4K           & 8K            & 16K          & 32K           & 64K           & 128K          & 256K          \\
$P_r \times P_c$ & $32 \times 64$ & $32 \times 128$ & $32 \times 256$  & $32 \times 512$ & $32 \times 1024$ & $32 \times 2048$ & $32 \times 4096$ & $32 \times 8192$ \\ 
\hline
\hline
     \textbf{16M}: $N_p/P$         & 8K      &  4K       & 2K       & 1K      & 512     & 256       & 128       &  64      \\
    $t_{comp}$ & 0.024 & 0.0123   & 0.0062 & 0.0032 & 0.0016 & 0.00089 & 0.00047 & 0.00023 \\
    $t_{comm}$ & 0.132  & 0.0974   & 0.0831 & 0.0408 & 0.0365 & 0.0297 & 0.0154 & 0.00902 \\
    total      & 0.156  & 0.114    & 0.0894 & 0.0441 & 0.0380 & 0.0307 & 0.0159 & 0.00928 \\
    \% comm    & 84.6\%  & 85.4\%  & 92.9\%  & 92.5\%  & 96.0\%  & 96.7\%  & 96.9\%  & 97.2\%  \\
    strong     & --      & 68.4\%  & 43.6\%  & --      & 58.0\%  & 35.9\%  & --      & 85.7\%  \\
    weak       & --      & --      & --      & --      & --      & --      & --      & --      \\ 
\hline
\hline
     \textbf{64M}: $N_p/P$        & 32K      &  16K       & 8K       & 4K      & 2K     & 1K       & 512       &  256      \\
    $t_{comp}$ & 0.093 & 0.049 & 0.025 & 0.013 & 0.0065 & 0.00035 & 0.0018 & 0.00095 \\
    $t_{comm}$ & 0.527 & 0.389 & 0.330 & 0.172 & 0.146 & 0.0122 & 0.0603 & 0.0379 \\
    total      & 0.621 & 0.440 & 0.354 & 0.185 & 0.153 & 0.126 & 0.0623 & 0.0389 \\
    \% comm    & 84.9\%  & 88.4\%  & 93.2\%  & 93.0\%  & 95.4\%  & 96.8\%  & 96.8\%  & 97.4\%  \\
    strong     & --      & 70.6\%  & 43.9\%  & --      & 60.5\%  & 36.7\%  & --      & 80.1\%  \\
    weak       & 100.5\% & 103.6\% & 101.0\% & 95.5\%  & 99.3\%  & 97.5\%  & 102.1\% & 95.4\%  \\ 
\hline
\hline
     \textbf{256M}: $N_p/P$        & 128K      &  64K       & 32K       & 16K      & 8K     & 4K       & 2K       &  1K      \\
    $t_{comp}$ & 0.380   & 0.198   & 0.099   & 0.0521  & 0.026  & 0.0143  & 0.0073 & 0.0038 \\
    $t_{comm}$ & 2.089   & 1.573   & 1.314   & 0.680   & 0.598   & 0.477   & 0.240   & 0.154 \\
    total      & 2.476   & 1.773   & 1.419   & 0.733   & 0.626   & 0.493   & 0.248   & 0.158 \\
    \% comm    & 84.4\%  & 88.7\%  & 92.6\%  & 92.8\%  & 95.5\%  & 96.8\%  & 96.8\%  & 97.4\%  \\
    strong     & --      & 69.8\%  & 43.6\%  & --      & 58.5\%  & 37.2\%  & --      & 78.4\%  \\
    weak       & 100.8\% & 102.9\% & 100.8\% & 96.3\%  & 97.1\%  & 99.6\%  & 102.6\% & 94.0\%  \\ 
\hline
\hline
    \end{tabular}
\caption{Timings for interpolation, computational + communication and total for different problem sizes. 
The three tables represent
particle counts of 16M, 64M and 256M. 
The weak scaling percentages are based on scaling up the number of particles
and reporting how much of the performance is retained.}
\label{tab:interp}
\end{table}

It can be seen that values of $t_{comm}$ in
Fig.~\ref{fig:break}(b) are higher than those for $t_{comp}$ in
Fig.~\ref{fig:break}(a), by an order of magnitude or more,
showing that communication is still dominant over computation.
This is consistent with trends in
Fig.~\ref{fig:break}(b) showing a strong
resemblance to those presented earlier in
Fig.~\ref{fig:interp.time} for the total interpolation time.
To analyze the communication timings more precisely
in the context of (\ref{eq:tcomm_a}) and (\ref{eq:tcomm_b})
we present some further details
including scalability percentages
in Table~\ref{tab:interp}.
For each choice of $N$, 
strong scaling is assessed with respect to the smallest
$P$ tested for each combination with $N_p$ also fixed, while
weak scaling is assessed relative to timings obtained with
$P$ increased in proportion to $N_p$.
It can be seen that while the strong scaling varies, 
weak scaling is close to perfect, which is consistent with 
observations that interpolation timings 
become almost proportional
to $N_p/P$.

%\textcolor{blue}{
%Recalling the discussion at the end of Sec.~3, we emphasize
%that the favorabe scaling characteristics seen in
%Fig.~\ref{fig:break}(b) is conditioned upon a low
%particle density $N_p/N^3$. If this ratio is raised
%substantially a cross-over point can be expected when
%the ghost layer approach for spline coefficients becomes
%favorable. The severe memory requirements of the ghost layers
%makes a precise performance comparison very difficult, since
%even the interpolation kernel itself is prone to running
%out of memory. However, some rough data suggests the cross-over
%point in $N_p/N^3$ is reached at about 64~M particles
%at $2048^3$, but as high as $O(10)$~G) at $8192^3$.
%}

Consistent with statements in the preceding paragraphs, the numbers
in Table~\ref{tab:interp} confirm that computation scales
almost perfectly, and is much less expensive than communication.
In this and the next paragraph we 
consider only data for the largest $N_p$ tested, i.e 256M.
For $N=2048$ $t_{comm}$  at 4K cores is
75.3\% of that at 2K cores, while
$t_{comm}$  at 8K cores is 83.5\% of that at 4K. This increase
indicates scalability is reduced. 
This can be partly understood using
(\ref{eq:tcomm_a}) with the ratios $(P_r+P_c)/P$ being
0.0469, 0.0391 and 0.0352 at 2K, 4K and 8K cores respectively: i.e.
decreasing less significantly between 4K and 8K.
However, for $N=4096$ the
scalability of communication time at 64K cores relative to
32K cores is better than that of 32K cores relative to 16K cores.
It is worth noting that at 64K cores the processor
grid used is such that $N/P_c=2$, which implies
(\ref{eq:tcomm_a}) ceases to be valid.
It can also be seen that for a given $N_p$, improved
scalability is generally obtained at larger $N$: e.g. 
$t_{comm}$ at 16K cores (with $N=4096$) is 
just 51.8\% (very close to 50\%)  of
that at 32K cores (with $N=2048$). A similar,
in fact slightly better  observation
can be made between  
$t_{comm}$ at 128K cores (with $N=8192$) and that
at 256K cores (with $N=4096$).

For the purpose of simulations at extreme problem sizes,
the most promising observation perhaps is that, for $N=8192$,
scalability between 128K and 256K cores 
is significantly better than those observed
at smaller problem sizes for a similar increase (doubling)
of core count. It is also remarkable that this has occurred
even though the communication time accounts for some 97\% of
the time taken for the interpolation. Clearly, this indicates
that the one-sided communication, achieved through Co-Array Fortran
is itself highly efficient and scalable
for our current problem configurations.

\begin{figure}[h]
\centerline{
\includegraphics[height=2.in]{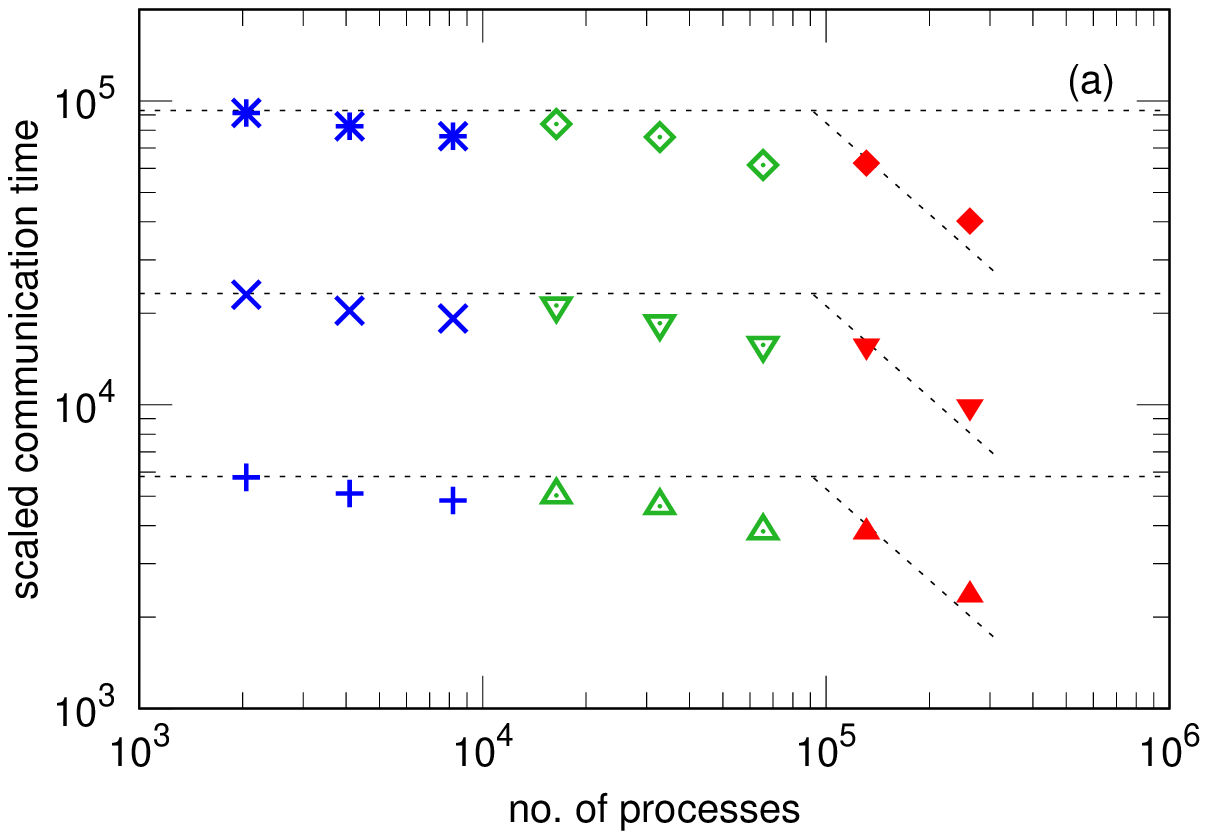}
\includegraphics[height=2.in]{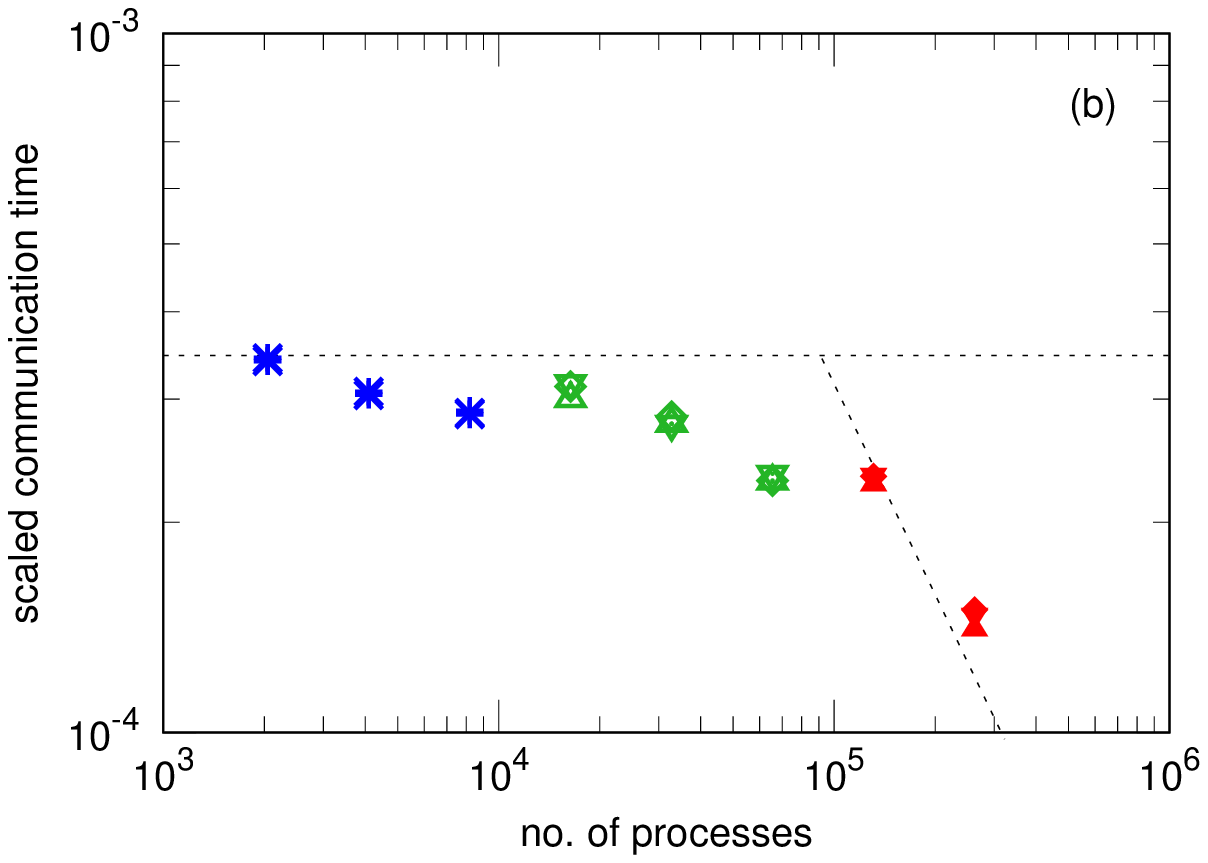}
}
\vspace{0.5mm}
\caption{
Scaled communication time $t_{comm}$ divided by
$(P_r + P_c)/(P \ N)$ in (a) and
further by $N_p$ in (b) . Same symbols as
Figure~\ref{fig:interp.time}.
The estimates 
(\ref{eq:tcomm_a}) 
and (\ref{eq:tcomm_b}) 
are represented by horizontal dashed lines and
lines of slope -1.
}
\label{fig:comm}
\end{figure}

For more precise tests of the scaling estimates
(\ref{eq:tcomm_a})
and (\ref{eq:tcomm_b})
we show in Fig.~\ref{fig:comm}
communication timings  (a) scaled by
the factor $(P_r+P_c)/(PN)$ and (b) further divided by $N_p$.
If $\Lambda$ is the proportionality factor in (\ref{eq:tcomm_a})
then the quantities being shown are $\Lambda N_p$ and $\Lambda$ itself
respectively. If (\ref{eq:tcomm_a}) holds perfectly then all
data points should lie on the same horizontal line.
The results suggest that (\ref{eq:tcomm_a}) holds quite
well at $2048^3$ but there is a gradual transition
to (\ref{eq:tcomm_b}) as grid resolution increases
to $4096^3$ and $8192^3$.  
As mentioned earlier, 
this transition is partly a consequence of our
practice (driven by considerations for FFT performance)
of keeping $P_r$ fixed as $P$ is increased,
which results in $P_c$ approaching and eventually becoming equal to $N$ at extreme problem sizes.
The strong downward
trend of these data points in the limit of large $P$
suggests good scalability of the new algorithm at
large core counts, as discussed above.
For each problem configuration the absolute value
of $\Lambda$ as in frame (b) of this figure may be considered
a measure of system communication performance  which
is inherently machine dependent.

The central characteristic of our new interpolation algorithm  is that,
by fetching only the spline coefficients for particles near the boundary,
communication requirements are reduced to a bare minimum.
This communication is also fast, being local (between adjacent MPI processes)
and highly expedited via PGAS techniques implemented via Co-Array Fortran.
In contrast, in the baseline algorithm before this work was conducted,
every MPI process has to  compute a partial sum for every particle, which involves 
a lot of wasteful calculations (with  most contributions being zero)
as well as communications. As a result,
there is a huge contrast between the timings for the two approaches. 
While the timings for the Eulerian based operations, such as FFTs and calculation
of spline coefficients are the same between two approaches, there
is more than an order of magnitude difference between the interpolation operations.
For the case of $N=8192$, $N_p=64$M and $P=256$K 
the original approach for interpolation (not counting the
spline coefficients) takes  about 18.5 seconds (Table~\ref{tab:global}),
but the new approach (Table~\ref{tab:interp})
takes only 0.0389 seconds.
%which translates to a speedup of greater than 400X.
Since the cost of the interpolation increases proportionally
with the number of particles, the difference in raw timings
will be  even greater when more particles are used.
The scalability analysis presented in this section also
suggests strongly that the new approach will continue
to perform well at even more extreme problem sizes of
$N$ and $N_p$ beyond the largest values tested in this work.
However, it is also important to note that $N_p$ always needs to be
sufficiently less than $N^3$ for the current algorithm to be most viable.
Since otherwise, if $N_p$ is comparable to $N^3$, 
the same spline coefficients would be redundantly fetched
multiple times for interpolation, making a ghost layer based approach
more viable.

\subsection{Particle migration}

The preceding discussion of communication performance
has focused on the transfer of spline coefficients,
which occurs for particles within a distance of up to 2 
grid spacings from a sub-domain boundary. Since
particles migrating to another sub-domain become
the responsibility of  a new host MPI process, we
also consider here the cost of communication involving
such particle migrations. 
At each Runge-Kutta sub-step, 
once the velocity of a particle is determined,
the particle position is updated at minimal computational cost.
Each MPI process then
scans the positions of all particles under its control,
identifying those which have just left its sub-domain. Because 
we use 2D decomposition, a migrating particle can potentially move
to one of $3^2-1=8$ possible adjacent sub-domains.
All attributes (including position and velocity information,
at minimum) associated with such departing particles  
are copied into a temporary outgoing buffer,
while the ordering of remaining particles is adjusted to 
fill gaps in array positions left by the departing particles.
Neighboring
MPI processes then exchange information on how many particles
are leaving from which MPI process to which MPI process.
The actual transfer of information is performed as a simple
halo exchange using non-blocking \texttt{MPI\_ISEND} and \texttt{MPI\_IRECV}
calls. Information on the arriving particles is then
appended to the array holding the incumbent particles. 
As noted earlier, because of homogeneity in space
for the turbulent flow, the number of
particles in each sub-domain is almost always close to $N_p/P$.

Clearly, the cost of handling particle migrations is largely
determined by the number of particles migrating. This number
is influenced by the Courant number ($C$) constraint on the
time step and the likelihood 
for particles to be located within $C$ grid spacings of
the sub-domain boundaries. For a given number of particles
per process and time step, the number of particles migrating
particles is also sensitive to the surface area to volume ratio
of each sub-domain. In particular if a sub-domain is
very thin in one direction (which happens
if $N/P_c$ is as low as 2 or 1) then considerable migration
activity in that direction is expected.

\begin{table}[h]
\begin{center}
    \begin{tabular}{c||ccc|ccc|cc}
    \hline
    \hline
    $N$      & 2048        & 2048         & 2048          & 4096         & 4096          & 4096          & 8192          & 8192          \\
    $P$       & 2K          & 4K           & 8K            & 16K          & 32K           & 64K           & 128K          & 256K          \\
    $P_r\times P_c$    & $32 \times 64$ & $32 \times 128$ & $32 \times 256$  & $32 \times 512$ & $32 \times 1024$ & $32 \times 2048$ & $32 \times 4096$ & $32 \times 8192$ \\ 
	$N/P_c$ &  32 & 16 & 8 & 8 & 4 & 2 & 2 & 1 \\
\hline
 \% migrating & 0.29\%         & 0.48\%    & 0.88\%       & 0.78\%      & 1.56\%       & 3.13\%       & 2.73\%       & 5.47\%        \\
    wall time & 0.0041       & 0.0026      & 0.0019       & 0.0022      & 0.0020       & 0.0019       & 0.0031       & 0.0142       \\
\hline
\hline
\end{tabular}
\end{center}
\caption{Timings (in seconds) for operations needed to
handle particle migration between different sub-domains using
\texttt{MPI\_ISEND} and \texttt{MPI\_RECV} calls  at different grid
resolutions and core counts, with number of particles held fixed at 256M.
}
\label{tab:mig}
\end{table}

Table \ref{tab:mig} shows elapsed wall time for
the particle migrations, averaged over all MPI processes
and over a large number of time steps in the DNS code.
A balance of competing effects leads to different 
trends as the core count is varied for several
grid resolutions.
For $2048^3$, with each doubling of core count
the number of particles per process decreases
by a factor of 2 while the ratio of surface area
to sub-domain increases by a factor less than 2, such that
the net result is a mild (less than 50\%) reduction in the time spent
in handling the migrations, as seen from 2K to 4K
and similarly from 4K to 8K cores.
For $4096^3$, as one side of each sub-domain 
becomes rather thin ($N/P_c$ falling from 8 downwards) the
two effects almost mutually cancel, as a greater fraction
of the $N_p/P$ particles now reside close enough to the
sub-domain boundaries for a migration to be possible.
Finally, at $8192^3$, as $N/P_c$ drops to 1, practically
every particle is in a zone where migration is possible,
i.e.  likelihood of migration for each particle is 
substantially increased, 
leading to a substantial increase in the migration timing.
It is likely that a less elongated shape of the 
processor grid, such as $64\times 4096$, with
no dimension being very thin, will result in
fewer migrations. However, since the time taken for migration
for the largest case in the table is still well under
1\% of the overall simulation time,
no special strategies for further optimization of this facet of
our algorithm appear to be necessary.

In regard to particle migration, we would also like to point
out that the general considerations in this sub-section
can change somewhat depending on the underlying flow physics or the type
of particles tracked. For example, in a compressible flow,
or if particle inertia is involved, local accumulations
contributing to greater load imbalance (i.e., 
greater departure from $N_p/P$ per MPI process) can occur.
In the case of inertial particles or molecular markers with
high diffusivity the particles can also move by
more than one grid spacing over one time step.
In such a scenario, the migration can be generalized by considering more 
sub-domains beyond the immediately adjacent ones, i.e., 
the halo-exchange is now extended to $(2n+1)^2 - 1$ possible
sub-domains (assuming 2D processor grid layout), 
where $n$ is the number of neighboring sub-domains 
on one given side. In these situations the present
algorithm is likely to be less efficient.
The development of 
more advanced coding strategies necessary to address
these further challenges is 
an interesting topic for future work.

\subsection{Summary of overall timings}

\begin{table}
\centering
    \begin{tabular}{c|cc|cc}
\hline
\hline
    $N$               & 4096          & 4096          & 8192          & 8192          \\
    $P$               & 32K           & 32K          & 256K           & 256K          \\
    $P_r \times P_c$ & $32 \times 1024$ & $32 \times 1024$ & $32 \times 8192$ & $32 \times 8192$ \\
    $N_p$            & 64M           & 256M           & 64M          & 256M          \\ \hline
    Eulerian        & 6.59          & 6.59          & 9.20          & 9.20          \\
    Splines         & 2.33          & 2.33          & 4.42          & 4.42          \\
    Particles       & 0.31          & 1.25          & 0.08          & 0.32          \\
    Total           & 9.23          & 10.17         & 13.70         & 13.94          \\
\hline
\hline
    \end{tabular}
\caption{Summary of total timings for the entire DNS code when using the dynamic particle-to-process
mapping.}
\label{tab:total}
\end{table}            

We close our performance analysis by showing in Table~\ref{tab:total}
the overall elapsed time per step for production simulations with
particle tracking based on the new algorithm developed in this paper.
It can be seen that the cost of interpolation is now primarily in the
calculation of cubic spline coefficients from the velocity field, while
the calculation of interpolated particle velocities from the spline
coefficients has become highly efficient, scaling mainly with the
number of particles per MPI process. 
The last case shown in this table shows the cost of
tracking 256M particles in our largest simulation is not much more than
50\% additional to the cost of computing the velocity field alone.
Although the cost of following
the particles does increase with $N_p$, we expect that most science questions
on Lagrangian statistics can be answered reasonably well using 
particle population sizes comparable to the largest values tested
in this work.

It may be recognized that there is an inevitable element of machine
and hardware dependency on the numbers presented here.
Clearly, our new algorithm is superior to the previous
static mapping based approach. However, we are unable to make
similar quantitative comparisons with other prevalent 
approaches based on utilizing ghost layers,
due to severe memory constrains associated with them
at extreme problem sizes. 
Nevertheless, 
depending on the machine hardware and interconnect along with the
number of particles tracked in comparison to the number of grid points,
it is possible that the ghost layer approach may
become more viable.
Further testing is still necessary for more definitive
claims on the overall relative merits of the Co-Array Fortran
and ghost layer approaches, especially on 
future machines with larger memory per node and
improved communication bandwidth.
However, since the new algorithm is based on the premise of
communicating as little and as locally as possible, the
general trends are likely to hold on other machines as well,
provided a robust PGAS-based programming model
is well supported, as it is on {\em Blue Waters}.

\section{Conclusions}

In this paper we have reported on the development of 
a new parallel algorithm for particle tracking in direct numerical
simulations (DNS) of turbulent flow, with the objective of addressing
challenges at extreme problem sizes, where a large number of fluid particles
are tracked at high grid resolution on a massively parallel computer.
The key task in following Lagrangian particle trajectories is to obtain
the particle velocity at its instantaneous 
position via interpolation from a set of fixed Eulerian
grid points.  Cubic spline interpolation is preferred because of
its high order of accuracy and differentiability \citep{YP.1988}.
From the velocity field on a 
periodic solution domain with  $N^3$ grid points distributed over
$P$ parallel processes using a 2D domain decomposition 
 (consisting of row and column communicators)
$(N+3)^3$ spline coefficients are first obtained by solving
tridiagonal systems in each coordinate direction. 
For each particle the interpolated velocity is obtained by 
summing over a stencil of $4^3=64$ spline coefficients which may
be distributed among different processes, thus requiring substantial
communication. Since the particles are free to wander under
the effects of turbulence, the interpolation
stencil for each particle also changes continually over each time step.
The existing algorithms in literature either distribute the 
particles among the processes in exactly the same manner at every time step
or use ghost layers to incorporate information from neighboring processes.
However, we find that these approaches 
are unable to provide acceptable performance at the largest problem sizes
currently known for Eulerian-only simulations.
% without fluid particles.

In this work, we have developed a new parallel algorithm where communication
is reduced to a bare minimum and occurs only 
between processes adjacent to each other in a 2D Cartesian 
processor grid, thus leading to very high scalability.
Particles are now distributed among the processes based on
their instantaneous position, and each process is responsible for 
a dynamically evolving group of particles,  such that all
interpolation information is available either locally on the host
process or its immediate neighbors. The most distinctive element
of our implementation is avoiding ghost layers (which leads to high memory
costs and wasteful communication) completely, in favor of 
one-sided communication through the use of a  partitioned global
address space (PGAS) programming model. In particular, we
use Co-Array Fortran (CAF), which is optimal for small messages
and is well-supported in the Cray Compiler Environment on
the petascale supercomputer {\em Blue Waters} operated by the
National Center for Supercomputing Applications (NCSA) at
the University of Illinois, Urbana-Champaign, USA.

In our new algorithm, the spline coefficients are stored
as a global co-array, which is logically partitioned into distinct sections
local to each process. 
At the beginning of each Runge-Kutta sub-step
each process decides, based on the proximity of each particle
to the sub-domain boundaries, which spline coefficients need to
be fetched from one or more of 8 neighboring processes.
The transfer is coded as a simple CAF assignment statement whose
execution is, because of local memory affinity and message
size being only 4 real words, extremely fast.
After this process is complete the host process performs
the summation of 64 coefficients and calculates the new
particle position. If a particle has moved to an adjacent
domain then a halo exchange is used to transfer its information
to a new host process. While these migrations occur on a regular basis
the fraction of particles migrating at any one time step is
generally low and hence the cost inconsequential.

Detailed benchmarking results obtained on {\em Blue Waters}
are reported, for problem sizes from $2048^3$ grid points
on 2048 Cray XE cores to $8192^3$ 
(over 0.5 trillion grid points) on 262,144 cores.
The calculation of spline coefficients is slightly less
efficient than 3D Fast Fourier Transforms (FFTs) which form
the backbone in our simulations of velocity fields and has
been optimized aggressively for our recent work \citep{YZS.2015}
performed using millions of node hours on {\em Blue Waters}.
With spline coefficients available, the cost of 
the remaining interpolation
operations is almost proportional to the number of particles ($N_p$)
but also sensitive to the Eulerian problem size ($N$) and the
shape ($P_r \times P_c = P$) of the 2D domain decomposition 
used for the latter. These particle costs
are dominated by communication which is especially sensitive
to the shortest dimension of each sub-domain compared to
the size of the interpolation stencil in each direction. Both
theoretical arguments and actual performance data (Figs.~4 and 5) indicate,
somewhat counter-intuitively, that the scalability of this
algorithm actually improves with increasing problem size.
For our $8192^3$ simulation with 256M particles
the performance improvement obtained
(see contrast between Tables 1 and 5) is so substantial that
the fraction of additional time for tracking 256M particles
compared with an Eulerian simulation is only about 50\%, which
is less than the corresponding fractional cost
for 16M particles in a $2048^3$ simulation.

In summary, in the work described in this paper we have
successfully overcome a challenge in 
scalability for tracking a large number of fluid particles
in a spatial solution domain which is distributed
over a large number of parallel processes.
The key is to aggressively minimize communication,
which is implemented using Co-Array Fortran (CAF).
based on a partitioned global address space (PGAS) 
programming model. Use of CAF is especially advantageous
in our application because the algorithm uses
small although numerous messages and is localized
between immediately neighboring processes.

Some of the conclusions of this paper may be partly
dependent on machine architecture, including the
network characteristics of {\em Blue Waters}.
However, since CAF is part of the Fortran 2008 standard,
the applicability of our CAF-based 
algorithm on other major platforms is expected 
to increase in the future.

\section*{Acknowledgments}

This research is part of the Blue Waters sustained-petascale computing project, 
which is supported by the National Science Foundation (NSF) awards OCI-0725070 and ACI-1238993
and the state of Illinois. Blue Waters is a joint effort of the 
University of Illinois at Urbana-Champaign and National Center for 
Supercomputing Applications (NCSA).
The authors gratefully acknowledge support from NSF,
via Grant ACI-1036070 (from the Petascale
Resource Allocations Program) which provided access to Blue Waters
and Grant CBET-1235906 (from the 
Fluid Dynamics Program) which provided the scientific impetus
for this research.
We thank Dr. R.A. Fiedler of
Cray Inc. for his advice on use of Co-Array Fortran,
staff members of the Blue Waters project for
their valuable assistance, and 
K. Ravikumar at Georgia Tech
for his help with the contents of Tables~1 and 4.
In addition, we are grateful to
anonymous referees for their constructive comments
which have helped improve the manuscript further.

\end{document}